\documentclass[12pt,preprint]{aastex} 
\usepackage{graphicx} 
\usepackage{natbib}

\newcommand{\skipthis}[1]{}

\def\nh3{$\rm{NH_3}$}
\def\NH3{$\rm{NH_3}$} 
 
\def\msun{M$_\odot$}
 
\def\lsun{L$_\odot$} 
\def\kms-1{km~s$^{-1}$}
\def\h2o{$\rm{H_2O}$} 
\def\h2{$\rm{H_2}$} 
\def\CM2{$\rm{cm^{-2}}$}
\def\cm3{$\rm{cm^{-3}}$} 
 
\def\vlsr{$\rm{V_{LSR}}$}
\def\n2h+{N$_2$H$^+$}
\def\ch3oh{CH$_3$OH}

\newcommand{\gsim}{${\raisebox{-.9ex}{$\stackrel{\textstyle>}{\sim}$}}$ }

\begin{document}

%\slugcomment{\emph{Draft \today}}

\title{IRDC G030.88+00.13: A Tale of Two Massive Clumps}

\author{Qizhou Zhang\altaffilmark{1}, Ke Wang\altaffilmark{1,2}}

\altaffiltext{1}{Harvard-Smithsonian Center for Astrophysics, 60 Garden Street,
Cambridge MA 02138, USA}

\altaffiltext{2}{Department of Astronomy, School of Physics, Peking University,
Beijing 100871, China}

\email{qzhang@cfa.harvard.edu}

%\keywords{clouds - - ISM: individual (G030+00.13) - ISM: kinematics and dynamics - stars: formation}

\begin{abstract}
Massive stars (M $\gsim 10$ \msun) form from collapse of parsec-scale molecular clumps. How 
molecular clumps fragment to give rise to massive stars in a cluster with a 
distribution of masses is unclear. We search for cold cores that may lead
to future formation of massive stars in a massive ($> 10^3$ \msun),
low luminosity ($4.6 \times 10^2$ \lsun)
infrared dark cloud (IRDC) G030.88+00.13. The \nh3 data from VLA and GBT reveal
that the extinction feature seen in the infrared consists of two distinctive clumps
along the same line of sight: The C1 clump at 97 \kms-1 coincides with the 
extinction in the Spitzer 8 and 24 $\mu$m. Therefore, it
is responsible for the majority of the IRDC. The C2 clump at 107 \kms-1 is more compact 
and has a peak temperature of 45 K. 
Compact dust cores and \h2O masers revealed in the SMA and VLA observations
are mostly associated with C2, and none is within the
IRDC in C1. The luminosity
indicates that neither the C1 nor C2 clump has yet to form massive protostars. But C1 might be
at a precluster forming stage. The simulated observations rule out 0.1pc
cold cores with masses above 8 \msun\ within the IRDC. The core masses in C1 and C2, and those in high-mass
protostellar objects suggest an evolutionary trend that the mass of cold cores increases
over time. Based on our findings, we propose
an empirical picture of massive star formation that protostellar cores
and the embedded protostars undergo simultaneous mass growth during the protostellar evolution.

\end{abstract}

\section{Introduction}

\skipthis{ Formation of OB clusters  \citep{lada2003} has been a 
subject of intense studies for several decades.}

It is well known that massive stars form in clusters that contain
a distribution of stellar masses \citep{lada2003}.
How dense molecular clouds contract and fragment to give rise to a cluster of stars
has been a subject of intense studies for decades. It is often suggested that the stellar 
mass at the peak of the Initial Stellar Function (IMF) is related to and perhaps is 
determined by the  characteristic mass in molecular clouds. The value of such mass 
is dictated by the interplay of gravity, internal pressure due to thermal  
and turbulent motions, and/or magnetic fields \citep{larson2005}. 
A typical proto cluster forming clump\footnote{We refer a molecular clump
as an entity with a size scale of 1 pc, which is capable of forming a cluster of stars, while a
molecular core as an entity with a size scale of 0.1 pc, which forms one or a small group of stars.} with massive young stellar objects 
contains $10^3$ \msun\ of dense molecular gas \citep{molinari2000,beuther2002a,beltran2006a}
within a size scale of about 1 pc. The average density ($10^5$ \cm3)
and temperature (15 K) \citep{rathborne2007, pillai2006b} in these clumps
yield a Jeans mass\footnote{Jeans mass $M_J = {\pi^{5/2} C_s^{3}\over 6 \sqrt{G^3\rho}}$, where $C_s$ is the sound
speed, and $\rho$ the mass density} of approximately 1 \msun. This characteristic mass 
explains naturally the peak of the IMF in a cluster \citep{larson2005}.

Massive stars, on the other hand, contain masses at least an order of magnitude 
larger than the global thermal Jean mass in a molecular clump. This presents
a theoretical puzzle, since cores and stars significantly
larger than the global Jeans mass are unstable
against fragmentations to less massive objects.  While the origin
of low mass stars could be linked to the Jeans mass, how massive stars form in a cluster
has been a matter of debate. Do massive stars form in 
cores of Jeans mass similar to their low mass counterpart, or form in
cores far more massive than the thermal Jeans mass? 
Observations of high-mass protostellar objects 
often find hot molecular cores with masses of $10 - 10^2$ \msun\ \citep{garay1999}, 
consistent with Jeans mass
at elevated gas temperatures of over 100 K.
However, the typical Jeans mass in a molecular clump
prior to massive star formation is much smaller
because of a lower temperature ($<$ 20 K) in these regions \citep{pillai2006b}.
In a sequence of evolution, cores that give rise to massive stars must be relatively
cold in the early stage due to a lack of strong stellar heating,  and later enter the hot core
phase characterized by  strong emission from
organic molecules as a result of evaporation of grain mantles and subsequent high 
temperature gas phase chemistry. 
In contrast to hot molecular cores with embedded high-mass protostars, we refer
entities that ultimately form massive stars,
$i.e.$, precursors of hot molecular cores, as {\it cold cores}.
What is typical mass of these cold cores? Do they contain $10^2$ \msun\ as postulated
in some theoretical models \citep{mckee2002}? Studies of infrared dark clouds (IRDCs), 
massive clouds at low temperatures and high densities \citep{egan1998, carey1998, carey2000, hennebelle2001, simon2006a, simon2006b, rathborne2006}, can provide clues
to these questions, and shed light on the process of massive star formation.

\citet{zhang2009} reported 1.3mm continuum image of
a massive molecular clump P1 in IRDC G028.34+00.06 (hereafter G28.34), obtained with the
Submillimeter Array. This clump contains more than $10^3$ \msun\ of dense gas, with a low
luminosity of a few 100 \lsun\ \citep{wang2008}.
Arcsecond resolution observations resolved
the molecular clump into five dense cores with masses of 22 to 64 \msun\, separated
by 0.14 to 0.2 pc. The gas temperature, measured at $3'' - 4''$ resolution
with the \nh3 (J, K) = (1,1) and (2,2) amounts
to 13 to 16 K in these cores \citep{wang2008}.  The average gas temperature and
density in the clump yield a thermal Jeans mass of 1 \msun,
1 to 2 orders of magnitude smaller than the mass of the cores. On the other hand,
the core masses are compatible with the turbulent Jeans mass\footnote{Turbulent Jeans mass follows
the same formula as the thermal Jeans mass ($M_J = {\pi^{5/2} C_s^{3}\over 6 \sqrt{G^3\rho}}$), 
but with the sound speed $C_s$ replaced by the turbulent velocity. Turbulence
can be highly anisotropic at large spatial scales. In the analysis, we assume isotropic turbulence
at the scale of dense cores.}, characterized by
the FWHM linewidth of 1.7 \kms-1. 

\skipthis{
In a similar context, \cite{longmore2010} examined 
fragmentation
in a more evolved massive molecular clump G8.68-0.36 with a luminosity of $10^4$ \lsun, 
and found cores between 3 to 10 \msun\, compatible with Jeans mass with elevated 
temperatures from heating by massive protostars.
}

The study of G28.34 is insightful in revealing the properties of cold cores. Here,
we present results from SMA, VLA and GBT of a massive
IRDC clump G030.88+00.13 (hereafter G30.88) with a mass and luminosity similar to G28.34.
The object, which is dark from 8 to 24 $\mu$m in the Spitzer images, 
contains a mass of $> 10^3$ \msun\ estimated from 850 $\mu$m continuum emission, embedded 
in a 20,000 \msun\ 
cloud seen in CO \citep{swift2009}.  The region, at a reported kinematic distance of 6.7 to 7.2 kpc,
has a luminosity of only 460 \lsun\ \citep{swift2009}. \citet{swift2009} presented arcsecond
resolution image at 870 $\mu$m with the SMA, which identified a massive 
core of 110 \msun.
The \nh3 data from the new VLA and GBT observations in this paper reveal two velocity components 
along the line of sight: One component coincides with the IRDC G30.88. The other component, 
likely at a different distance, is associated with the massive core detected
with the SMA and an \h2O maser emission. No compact cores more massive than 8 \msun\ ($4 \sigma$) 
are found to be
associated with the IRDC clump. Thus, this IRDC component may represent a precluster clump.
In Section 2 we presents details of the observations. Section 3 presents the main results and
the implications of the findings to cold core formation and evolution.

\section{Observations}

\subsection{VLA}
We observed IRDC G30.88 in the \NH3 (J, K) = (1,1) and the
(2,2) lines with the VLA\footnote{The
National Radio Astronomy Observatory is operated by Associated
Universities, Inc., under cooperative agreement with the National
Science Foundation.} on 2010 January 8 in its D configuration. The
phase center of the observations was RA(J2000) = 18:47:13.70,
Dec (J2000) = $-$01:45:03.70. The 25m-dish of the VLA antennas
yields a FWHM primary beam of approximately 2$'$ at the observing frequencies. 
We employed the 2IF mode that splits the
256-channel correlator into two sections to observe 
the \NH3 (1,1) and (2,2) lines simultaneously in one
polarization for each line. The channel separation was 24.4
KHz ($\sim$0.3 km s$^{-1}$ at the line frequencies). The time
variation of antenna gains was calibrated using QSO J1851+005,
observed at a cycle of about 20 mins. The absolute flux density is
established by bootstrapping to 3C286. The bandpass is calibrated
via observations of 3C454.3.

Subsequently, we observed the \NH3 (J, K) = (3,3) line with the EVLA on 2010 May 09
in its D configuration. We observed two polarizations using a bandwidth
of 4 MHz that splits into 256 channels. The time
variation of antenna gains was calibrated using QSO J1851+005,
observed at a cycle of about 6 mins. The absolute flux density is
established by bootstrapping to 3C48. The bandpass is calibrated
via observations of 3C454.3.

Data on the 22 GHz \h2O maser transition were obtained with the EVLA on 2010 December 16
in its C configuration. We observed two polarizations using a bandwidth
of 4 MHz that splits into 64 channels. The time
variation of antenna gains was calibrated using QSO J1851+005,
observed at a cycle of about 7 mins. The absolute flux density is
established by bootstrapping to 3C48. The bandpass is calibrated
via observations of 3C454.3.

The visibilities were calibrated and imaged using the AIPS/CASA
software packages of the NRAO. 
The synthesis beam is about $4''\times 3''$ for the D array data, and $2''\times 1''$ 
for the C array data. The detailed observational parameters are summarized in Table 1.

\subsection{GBT}

Observations of G30.88 were carried out with the Green Bank Telescope (GBT)$^1$ 
in the \nh3 (J,K) = (1,1), (2,2), (3,3), and (4,4) transitions during 2010 February 27 
through March 1, and 2010 April 29. We used the K-band receiver and a spectrometer
setup in the frequency switching mode that covers four windows, each with 12.5 MHz 
bandwidth split into 4096 channels. System temperatures varied
from 50 to 100 K during the observations mostly due to changes in elevation. The 100m
aperture of the GBT gives a $30''$  FWHM in the primary beam. Data 
were processed using GBTidl. We only present data of the (3,3) line in this paper.

\subsection{SMA}
Observations with the SMA\footnote{The Submillimeter 
Array is a joint project between the Smithsonian
Astrophysical Observatory and the Academia Sinica Institute of Astronomy and
Astrophysics, and is funded by the Smithsonian Institution and the Academia
Sinica.} \citep{ho2004}  were made with 8 antennas
in the compact configuration from 2007 July  7 through 2008 June 15, in a mosaic
of 18 pointings by \citet{swift2009}.
The zenith opacity at 225 GHz were 0.08 to 0.15
for the four nights, with double sideband system temperatures of
400 during the transit. 
The receivers were tuned to an LO frequency of 351 GHz for the 2007 observations,
and 340 GHz for the 2008 observations. Table 1 summarizes the main observational parameters.

The visibility
data were calibrated with the IDL superset MIR package developed for the
Owens Valley Interferometer.  The absolute flux level is accurate to about 25\%.
After
calibrations in MIR, the visibility data were exported to the MIRIAD format for
further processing and imaging.  The continuum is constructed from line
free channels in the visibility domain. 
We combined the continuum data from four tracks, which yields
a $1\sigma$ rms of 5.7 mJy and a synthesized beam of about 1.9$''$
with the naturally weighting of the visibilities.

\section{Results and Discussions}

\subsection{Dense Molecular Gas and Dust Continuum}

Figure 1 shows an overview of the G30.88 region in 24 $\mu$m obtained from the 
Spitzer MIPSGAL \citep{carey2009},
overlaid with contours of the \nh3 (1,1) emission from the VLA. The 850  $\mu$m 
continuum from the JCMT archive (see also \citealt{swift2009}) is also outlined in the yellow
contour. The \h2O masers detected with the VLA are marked with the plus sign, and
the continuum sources detected with the SMA are marked by the star symbol.
At 24 $\mu$m, G30.88 is dark against the galactic IR background. The IR extinction
matched the \nh3 and the 850 $\mu$m continuum emission well.

Figure 2 presents the integrated \nh3 (J,K)= (1,1), (2,2) and (3,3) emission from the VLA, 
the 850 $\mu$m continuum from JCMT, and 870 $\mu$m continuum
from the SMA. The range of velocities for the integrated emission is 
from 95 through 110 \kms-1. The \nh3 data reveal a dense gas filament that follows 
the dust emission from JCMT (approximately $14''$ resolution). It appears that the
\nh3 emission is spatially extended in the (1,1) transition, 
and becomes far more compact in the (3,3) line. Since
the \nh3 emission of the (1,1), (2,2) and (3,3) arises from energy levels of 23K, 65K 
and 125 K, respectively, a progressively smaller spatial extent in higher excitation lines
indicates that the extended \nh3 gas is at  relatively lower temperatures.

The SMA observations resolved the dust emission seen at lower resolution into
a dominant compact feature as reported in
\citet{swift2009}. This source is slightly resolved  at a resolution
of $1''.9 \times 1''.8$, and consists of two peaks. The maximum and the integrated 
flux of the source is 119 mJy and 255 mJy, respectively. 
In addition to the dominant core which we name SMM1, there appear to
be emission peaks at a level of 4 to 8 $\sigma$ rms. 
We identify all dust peaks above 5  $\sigma$ rms. Following
SMM1, we name other five peaks SMM2 through SMM6, ordered in decreasing fluxes.
The parameters of the continuum sources
are given in Table 2. Of the six sources, only SMM1 was reported in \citet{swift2009}.
With a more accurate temperature measurement (45K, see Section 3.2) for SMM1, we find a gas mass of 32 \msun,
about 1/3 of the value in \citet{swift2009}. The difference is mainly due to the temperature
values used. By comparing the dust peaks with the \nh3 and \h2O data, we confirm that
all the dust peaks are robust detections (see Section 3.4).

\skipthis{
We ignore
the peaks close to the edge of the mosaic since the primary beam correction amplifies
the noise level toward the edge of the map. The parameters of the continuum sources
are listed in Table 1.}

Figure 3 presents the VLA spectra of the \nh3 (1,1), (2,2) and (3,3)
transitions toward SMM1. For comparison, we also present the \nh3 (3,3) spectrum
from the GBT for the same position. The \nh3 metastable lines have 18 distinctive hyperfines
that normally appear in 5 separate components due to blending \citep{ho1983}.
For the (1,1) transition, the inner and outer satellite pairs appear
at 7.7 and 19.4 \kms-1\ from the main hyperfine. For the (2,2) transition, the inner
satellite pair is 16.6 \kms-1\ from the main hyperfine. Both the (1,1) and (2,2) spectra
show complex features indicative of multiple velocity components. The (3,3) line, for which the inner 
hyperfines are further out  from the main line and are much
fainter, clearly reveals two line-of-sight velocity components,
C1 at a \vlsr\ velocity of 97 \kms-1, and C2 at 107 \kms-1. 
The relative strength in brightness temperatures between the VLA and GBT spectra
indicates that the C2 component is spatially compact, thus is much brighter at
a higher angular resolution (0.35 K in GBT versus 5.0 K in the
VLA map). On the other hand, the C1 component is more extended: It is
detected by the GBT at a peak temperature of 0.2 K, and is not seen in the VLA spectrum
at a $1 \sigma$ rms of 0.5 K. A comparison between the GBT flux and the $1 \sigma$
rms of the VLA data suggests that the faint (3,3) emission seen in C1 is spatially
extended and fills the GBT beam.

\subsection{Properties of Two Cloud Components}

Due to the crowded hyperfine structures in
the (J,K) = (1,1) line, the main hyperfine of the C1 component is
blended with the first satellite hyperfine of the C2 component.
Since the hyperfines of the (2,2) line are further separated, we compute
its moment 0, 1, and 2 maps for the two velocity components, which are
presented in Figure 4. As one can see, the
morphology of the C1 component is more extended, and 
correlates well with the
extinction feature in the 24$\mu$m image and sub-millimeter continuum emission 
from the JCMT. The C2 component is spatially compact and is more centrally 
peaked than the C1 component. An \nh3 extension to the northeast of SMM1 seen in the C2 component
coincides with the IR extinction and probably contributes to the IRDC. However, the integrated
flux of the extension amounts only 20\% of the integrated flux over the same area
in the C1 component. Therefore, IRDC is predominantly associated with the C1 component.
SMM1, on the other hand, coincides with the peak of 
the \nh3 emission of the C2 component, and falls in the trough in the \nh3 emission 
of the C1 component.
Thus, SMM1 is likely associated with the 107 \kms-1\ C2 component. We will further 
discuss the association of other continuum sources with the two cloud components in Section 3.4.

The LSR velocities of the dense \nh3 gas
yield a kinematic distance of 6.5 and 7.3 kpc for the two cloud components, respectively.
The integrated fluxes of the \nh3 (2,2) emission within the IR extinction are 1.9
and 1.0 Jy \kms-1\ for the 97 and 107 \kms-1\ components, respectively. 
The JCMT 850 $\mu$m flux integrated within the IRDC region amounts to 9 Jy. 
Assuming that
the dust continuum emission from the two components is proportional to
the fluxes of the corresponding \nh3 emission, we obtain an 850 $\mu$m flux
of 5.9 Jy for C1, and 3.1 Jy for C2, respectively.
Using an average dust temperature of
19 K (see temperature estimate later in this section), a dust opacity 
law of \citet{hildebrand1983} with a
spectral index of 1.5, and a dust to gas ratio of 1:100, we obtain a mass
of $1.8 \times 10^3$ \msun\ for C1, and $1.2 \times 10^3$ \msun\ for C2. 
With a size of $40'' \times 26''$ for C1, the average 
density amounts to $2.0 \times 10^4$ \cm3 and $1.0 \times 10^{23}$ \CM2, respectively.
The dust opacity law adopted here gives an opacity value $\kappa (850 \mu m) = 0.015$ \CM2~g$^{-1}$,
consistent with that used in \citet{swift2009}.
Adopting a spectral index of 2 for the dust opacity law
results in a nearly factor of 2 larger in gas mass.

We estimate the \nh3 gas temperature \citep{ho1983} using the VLA data. Since the
hyperfines of the C1 component is blended, we cannot reliably obtain its optical depth. 
Assuming optically thin emission for the C1 component,
we derive a rotational temperature of 20 K. The rotational temperature $T_R(2,2:1,1)$
is related to the flux ratio of the main hyperfine component F(1,1) and F(2,2)
(see eqn 4 in \citealt{ho1983}) via
$$ T_R(2,2:1,1) = -41.5 \div ln\{ {-0.283 \over{\tau_m(1,1)}} ln [1 - {F(2,2)\over{F(1,1)}} \times (1 - e^{-\tau_m(1,1)})]\}.$$
When $F(2,2) < F (1,1)$, which is the case for C1, the optically thin 
assumption causes the rotation temperature 
to be over estimated. For $\tau_m(1,1)$ = 1 and 5, the gas temperature of 20K under
the optically thin assumption becomes
19 and 12 K, respectively. The \nh3 (1,1) emission likely has a moderate optical depth, 
thus the gas temperature of the C1 component should be lower than 20 K.
For the C2 component at 107 \kms-1, the \nh3 emission is detected in the (3,3) line toward SMM1.
We find a rotation temperature of 45K toward SMM1, and 19 K outside of SMM1.

As shown in Figure 4, the line width of the C1 component revealed by
the moment two map is rather uniform, with an average line width of
0.55 \kms-1\ (or a FWHM of 1.4 \kms-1 \footnote{If the line profile is Gaussian, 
the FWHM linewidth is related to the Doppler linewidth by a multiplicative factor 
of $2\sqrt{2 ln2}$.}). There appears to be two areas with slightly larger
line widths of $>$ 1 \kms-1. The one northeast of SMM1 is due to broadening by a feature
at a slightly different velocity as shown in the moment 1 image. The other one 11$''$ south
of SMM1 has a line width of 2 \kms-1. This region coincides with a tentative detection
of an SMA peak at RA(J2000) = 18:47:13.46, Dec(J2000) = $-$01:45:13.0 
and with a peak flux of 27 mJy. Since the flux is lower than the $5 \sigma$ cutoff limit, we
do not report the source in Table 2. Future deeper observations will help
confirming this dust peak.

The Doppler line width of the C2 component varies from 0.34 (or FWHM of 0.8 \kms-1)
to 1.3 \kms-1\ (or FWHM of 2.5 \kms-1) toward the SMM1 dust peak. Both C1 and C2
components show a velocity  gradient. There is a velocity shift of
approximately 3 \kms-1\ over a projected length of
$40''$ (or 1.7 pc) along the filament in the
northeast-southwest direction in C1. For C2, there appears to be a shift of 1 \kms-1\
across SMM1 at a position angle of 60$^\circ$.  This motion is likely due to rotation
in the core similar to that found in high-mass protostellar objects (e.g.
IRAS 20126+4104 \citealt{zhang1998b, keto2010}, G10.6 \citealt{keto1987a}).

\skipthis{
The \nh3 (3,3) emission is clearly detected with the GBT toward the 
position of SMM1. As shown in Figure 3, the brightness temperatures of
the (3,3) line are 0.19 K and 0.33 K 
for C1 and C2, respectively (using a beam efficient of 67\% for this frequency band). 
The (4,4) line is marginally detected only for the 107 \kms-1\
component at a brightness temperature of 0.04 K. Using these values,
and assuming the same filling factor for the two transitions, we derive
a gas temperature of 40 K for the C2 component, and
an upper limit of 40 K for the C1 component. With a beam of $30''$,
we cannot constrain the spatial extent of the the warm gas. Future high resolution 
observations with the VLA will help revealing the spatial distribution of the warm gas.
}

\subsection{H$_2$O Masers}

As signposts of star formation, \h2O masers trace protostellar activities of a wide
range of stellar masses \citep{wouterloot1986, churchwell1990, palla1993,
claussen1996}. Since their excitation requires high density and 
temperature \citep{elitzur1989, felli1992}, \h2O masers normally arise in
the close proximity of a protostar, thus mark the position of the protostar that may
not be revealed otherwise. A total of seven \h2O masers are detected in the
region. Table 3 reports the maser positions and the peak brightness temperatures in
a descending order. Figure 5 presents the maser spectra. The \h2O maser fluxes reported are not
corrected for the primary beam of the VLA. This is because outside the FWHM of the
primary beam ($\sim 120''$), a symmetric two dimensional Gaussian profile does
not represent the beam accurately. Applying a Gaussian correction to the data
may introduce additional error in the maser brightness temperature, especially for masers outside of the
FWHM of the primary beam. Should such a correction be applied to the data,
the maser feature {\it{2}}, the one furthest away from the pointing center, is about
a factor of 5 brighter.

As seen in Figure 5, all masers exhibit complex spectra and have a
broad linewidth of 3 to 8 \kms-1. Since a typical maser feature
has a line width of $\sim$ 1 \kms-1\ as required by coherent amplification, the broad
linewidths seen here indicate spatially unresolved maser features in the synthesized beam
of $2'' \times 1''$. Masers {\it{1, 2, 6}} and {\it{7}} have centroid velocities close to 107 \kms-1.
Therefore, they are likely associated with the C2 cloud component (see more discussions
in Section 3.4). On the other hand, maser {\it{4}} has a centroid velocity 
close to 97 \kms-1. Thus, it is likely associated
with the 97 \kms-1\ component. We caution that masers are often excited in protostellar
outflows, and may have velocities offset from that of the cloud. This is shown in
masers {\it 3,} and {\it 5}, who have features detected close to 97 and 107 \kms-1.
We will discuss associations
of the dust continuum sources and \h2O masers with the thermal \nh3 gas in Section 3.4.

Other than SMM1 and SMM5, no other SMA dust peaks coincide with
\h2O masers. On the other hand, most \h2O masers are associated with
either \nh3 peaks, mid-IR sources or with the JCMT 850 $\mu$m continuum emission.

\subsection{Association of SMA Dust Peaks and \h2O Masers with the \nh3 Gas}

The continuum image from the SMA reveals at least six dust peaks with masses of
18 to 40 \msun. 
Except spatially extended CO, no other molecular line emissions are detected toward most of 
the dust peaks despite a total passband of 4 GHz. Thus, we cannot determine the \vlsr\ velocity
of these continuum sources based on the SMA data. In this section, we compare the spatial 
positions
of the SMA dust peaks with the \nh3 (1,1) data and the \h2O positions from the VLA to 
determine the association of the sources with the two cloud components. 
Figure 6 presents the \nh3 (1,1) channel maps of the
two cloud components with dust continuum peaks and \h2O maser positions.
It is clear that SMM1, the dominant dust peak, coincides with the peak \nh3 emission
(see velocity channels with \vlsr\ of 106.3 through 106.9 \kms-1\ and 109.1 through 109.7
\kms-1\ in Figure 6b). In addition, the peak velocity of the maser feature {\it{1}} 
is also close to the cloud velocity of component C2. Therefore, SMM1 and maser {\it 1} 
are associated with the C2 component.
The dust cores SMM2, SMM3, SMM4 and maser {\it{3}} are aligned in the north-south direction,
and coincide with an \nh3 filament (see channels with \vlsr\ of 106.9 through 109.4 \kms-1\
in Figure 6b). Thus, they are likely associated with the C2 cloud component.
SMM5 and maser {\it 4} are not close in projection to any \nh3 emission in the C2 component,
but is in close proximity of the extended \nh3 emission in the velocity channel of 95.8 \kms-1. Since
SMM5/maser {\it 4 } lie outside of the primary beam of the VLA, the extended \nh3 emission
is attenuated. Given that the maser velocity is close to that of the C1 cloud, we suggest
that SMM5/maser {\it{4}} are associated with the C1 component. Finally, SMM6 coincides with
the \nh3 emission in \vlsr\ of 106.3 through 107.2 \kms-1\, and 97.3 through 98.2 \kms-1.
It appears that the \nh3 (1,1) emissions in 106.3 through 107.2 \kms-1\ are
more centrally peaks. Therefore, we tentatively assign SMM6 to the C2 cloud component.
No strong \nh3 emissions are detected toward masers {\it 6} and {\it 7}. The peak velocities
of the masers, however, indicate that they are associated with the C2 component.
Tables 2 and 3 summarize the association of the SMA dust peaks and \h2O masers with
the two cloud components.

\subsection{Nature of the C2 Clump: A Massive Protocluster in the Making}

It appears that the majority of the SMA dust peaks (SMM1, SMM2, SMM3, SMM4, SMM6),
and \h2O masers ({\it{1,2,3,6,7}}) are associated with the C2 cloud component.
The presence of \h2O maser emission indicates protostellar
activities. The dominant dust continuum source, SMM1, is the only continuum
source associated with strong \nh3 (3,3) emission from
the warm gas of 45 K. Other continuum sources are not detected in
\nh3 (3,3) and are associated with gas of about 19 K.

Assuming that the dust is in a thermodynamic equilibrium with the gas in this high
density environment, we approximate the dust temperature by the \nh3 gas temperature.
For a dust temperature of 45 K and a source size of $1.1''$ for SMM1, we estimate a 
luminosity of 420 \lsun\ for SMM1 following \citet{scoville1976}. 
This value is consistent with the luminosity of 460 \lsun\
in \citet{swift2009}, derived from the spectral energy distribution. 
These luminosities correspond to a zero age main sequence star 
of 5 \msun. Since the star is under active accretion, which contributes to the 
total luminosity, the stellar mass is likely smaller.
This young protostar is surrounded by 32 \msun\ of dense gas 
within a scale of 0.04 pc, and several $10^3$ \msun\ gas on larger scales. There is
a large reservoir of dense gas for the protostar to accrete. The amount of dense gas and
the size scale of the molecular clump are similar to those that harbor massive stars
with bright hypercompact and ultra compact HII regions. For instance, one of the nearest examples of massive star formation, Orion KL, contains a dense molecular clump of several 
$10^3$ \msun\ within a scale of 1.5 pc \citep{chini1997,lis1998,johnstone1999}. 
This region, with a luminosity of $10^5$ \lsun\, 
is forming a cluster of stars \citep{beuther2005a, zapata2009}. As a similar example but at a larger
distance of 6 kpc, G10.6 is
a bright hyper compact HII region embedded in a flattened molecular clump 
of $> 10^3$ \msun\ in its inner 0.5pc region \citep{keto1987a,sollins2005a,liu2010a}. 
The region has a luminosity 
of $9 \times 10^5$ \lsun, and contains a cluster of stars of 195 \msun\ \citep{sollins2005a}
based on its ionization flux and total luminosity. The presence of several massive cores and 
\h2O masers in G30.88 C2 indicates active ongoing star formation.  The similar
large amount of dense gas and the low luminosity suggest that the G30.88 C2 region is a
younger cousin of Orion KL and G10.6, with the most massive protostar(s) still at 
an intermediate mass stage. It is reasonable to expect that it will form a massive cluster
when accretion is complete.

\subsection{Nature of the C1 Clump: A Massive Precluster Clump ?}

Beside the tentative detection of the dust peak at RA(J2000) = 18:47:13.46, 
Dec(J2000) = $-$01:45:13.0, there appear to be no dust continuum sources nor \h2O masers 
associated with the C1 cloud component within the IR dark region. 
Since the C1 cloud component contains $1.8 \times 10^3$ \msun\ over a scale of 2 pc, similar to
that of C2, a lack of compact cores
is intriguing. The relatively flat distribution of the \nh3 gas shown in Figure 4
and Figure 6a suggests that C1 is likely at an earlier evolutionary stage
than C2, which makes it a candidate for probing the onset of massive star formation.

It is often debated whether massive stars form from the monolithic collapse of cold,
dense molecular cores of $10^2$ \msun\ (e.g. \citealt{krumholz2005a}), or they form
as an integral part of a cluster formation in a $10^3$ \msun\ clump (e.g. 
\citealt{bonnell2004, li2004}). A key difference between the two pictures is whether
cold cores forming massive stars acquire all the mass initially before the birth of
a protostar.
It appears that the C1 clump bears the closest resemblance in conditions to the onset
of a massive star formation. In this clump, we find an upper limit of temperature of 20K, an
average density of $2.0 \times 10^4$ \cm3, and a line width of 0.55 \kms-1\ (or FWHM 
of 1.4 \kms-1). There appear to be no massive cold cores of $10^2$ \msun\
detected in the C1 clump, 
in contrast to what has be assumed in the monolithic collapse model. 

However, interferometers filter out spatially extended emission. To test possible
bias in the SMA observations, we examine the filtering effect and the ability in detecting
faint cores by simulating observations
using the JCMT image presented in Figure 2. We choose a single SMA pointing
centered at RA(J2000) = 18:47:12.70, Dec(J2000) 
= $-$1:45:22.90 for the simulation. A single pointing is preferred over
simulating multi-field mosaic in order to avoid
amplifications of noise toward the edge of the primary beam in mosaic. 
This particular pointing, which lies in between SMM1 and SMM5, is chosen since 
it is more than one primary beam away from any dust peaks.
Thus,  the image is less affected by the side lobes from SMM1. Figure 7 presents
a comparison of the SMA data with simulated images. The dust emission model,
shown in Figure 7b, is derived from the JCMT image
tapered by the SMA primary beam, approximately $34''$ at 345 GHz.
The peak flux of the JCMT data is 1.37 Jy at a beam of $14''$. Assuming that the emission
is smooth within the $14''$ beam, we derive a flux of 24 mJy per $1''.9 \times 1''.8$ beam.
This is about the 4$\sigma$ noise level of the SMA observations. Therefore, 
the continuum emission would have been barely detectable with the SMA, if it is smooth
and if there was no spatial filtering effect of the interferometer. A simulated observation
using the JCMT data as the source model, and using the UV coverage and noise characteristics
of the actual SMA data is shown in Figure 7c. As one can see, nearly no emission is 
detected due to a missing flux.
However, if there were a compact core with a flux of
30 mJy embedded in the extended molecular clump as a result of core formation, 
SMA observations would have reliably detected
such an object, as shown in Figure 7d. The fact that SMA does not
detect them limits the presence of compact cores of 8 \msun\ at a 4$\sigma$ level.

These simulated observations reinforce identification of the dust cores in
the SMA image reported in Table 2. 
In addition, the simulation indicates that at very early stages of massive star formation,
no massive cold cores {are spatially distinct from} the molecular clump. Our study of G30.88 rules 
out the presence
of 0.1 pc cores of 8 \msun\ at a $4\sigma$ level in the C1 region. 
This implies that cores giving birth to massive stars may not be initially 
massive (e.g. $10^2$ \msun).

\subsection{Fragmentation and Massive Star Formation}

The core masses detected in the G30.88 C2 clump range from 18 to 40 \msun. They are
likely the sites to form massive stars. A density of $2.0 \times 10^4$ \cm3
and a temperature of 19 K yield a thermal Jeans mass of 5 \msun, and a Jeans 
length of 0.2 pc for the region. Although the SMA observations are not deep enough to
detect cores of one Jeans mass, the cores revealed here are 4 times or more
massive than the Jeans mass, similar to the case seen in the G28.34-P1 clump \citep{zhang2009},
and G8.68 \citep{longmore2010}.
Assuming an isotropic turbulence for which its velocity is characterized by the \nh3
line width of 0.55 \kms-1, we find a turbulent Jeans mass of 37 \msun, a
value more compatible with the masses detected in the region. This reinforces the
notion that massive cores arise from a turbulent supported fragmentation. 

Similar to C2, the C1 region contains a large mass of $1.8 \times 10^3$ \msun. However,
in contrast to strong dust cores revealed by the SMA, no apparent compact cores
are detected at a $4 \sigma$ mass of 8 \msun. Assuming that C1 and C2 represent 
two stages along a common evolutionary sequence,
this implies that cold cores forming massive stars are probably less massive
than 8 \msun\ initially.
These cores will continue to grow in mass by gathering material from the clump,
and become compact cores as seen in C2.

An empirical picture of massive star formation appears to emerge from observations in 
the past several years. Surveys of cluster forming clumps reveal
blue-red asymmetry in optically thick tracers (e.g. HCN and HCO$^+$,
\citealt{wu2003, fuller2005, wu2007}), consistent with
infall of $1 - 2$ \kms-1\ at the cluster forming scale of approximately 1 pc.
These surveys with single dish telescopes at $> 10''$ resolution are in
agreement
with high resolution work in G10.6, in which global infall is observed in both
molecular and ionized gas toward a cluster of stars with a total stellar mass
of 195 \msun \citep{ho1986, keto1987a, sollins2005a, liu2010a}. Recently,
\citet{galvan2009} reported studies of an HII region G20.08-0.14,
where inverse P-Cygni profiles are detected toward a cluster of HII regions with
a linear size of 0.3 pc.
The infall motion appears to continue toward a hot molecular core in which
the youngest hyper compact HII region is embedded. 
The wide spread infall motion, together with the fragmentation studies in this
paper and \citet{zhang2009} suggests a picture of massive star formation in a
cluster forming environment: Collapse of $10^3$ \msun\ molecular clumps
as a result of losing internal turbulent support leads to the
formation of dense molecular cores. Among the cores, those leading to the birth of 
massive stars are more massive than the thermal Jeans mass. How these clumps
fragment to super Jeans masses is not certain, but turbulence in these cores
appears to be sufficient to provide the support \citep{zhang2009}.
During the early evolution, cores continue to draw material from the
molecular clump while the protostar embedded undergoes accretion. This picture is 
somewhat similar to
the competitive accretion by \citet{bonnell2004}. However, it differs in two
important aspects: First dense cores harboring massive stars are more massive
than the thermal Jeans mass; and secondly, accretion is likely dominated by
gas accretion in response to gravity as suggested in \citet{wang2010}, rather
than Bondi-Hoyle accretion. This empirical picture is derived from limited
observations. ALMA will deliver orders of magnitude improvement in continuum and
spectral line sensitivity as compared to current (sub)mm interferometers. More
importantly, large number of antennas will improve the dynamic range, and allow
detection of lower mass cores in the vicinity of bright objects.
It is hopeful that more sensitive studies of cold and massive
molecular clumps in the future will lead to a clearer and more complete picture of 
cluster formation.

\section {Conclusion}
In conclusion, we present spectral line and continuum images of a massive IRDC
G30.88. The cloud appears to consist of two line-of-sight
components C1 and C2 with LSR velocities of
97 and 107 \kms-1, respectively. Both molecular clumps are massive enough ($> 10^3$ \msun)
to form massive stars, but only C2 exhibits protostellar activities. Among the seven \h2O masers 
detected with the VLA, five have velocities or positions
associated with C2, and none is within the IRDC in C1. The SMA observations 
reveal six dust features
SMM1 through SMM6 with masses from 18 to 40 \msun,  much more massive than the thermal Jeans mass. 
Among the six cores, five are associated with 
the C2 region, and one is associated with the C1 component away from the main
extinction region. The \h2O maser emission and dust peaks in the C2 clump indicate active star formation,
but the low luminosity constrains the protostar(s) at an intermediate mass stage.
A lack of dust peaks and \h2O maser emission in C1 puts the IRDC at an even earlier stage 
of star formation, and does not support the idea of cold cores of $10^2$ \msun.
Observations of G30.88 and other IRDCs such as G28.34 seem to point to the early
evolution of massive star formation, in which cores gain mass from the clump while 
protostars accrete gas from the core.

\acknowledgements
We thank Jonathan Swift for permission of using the SMA data, and the anonymous
referee for comments that improve the clarity of the paper. Q. Z. acknowledges the
support from the Smithsonian Institution Endowment Funds. K. W.  acknowledges the
support from the SMA predoctoral fellowship and the China Scholarship Council.

{\it Facility:} \facility{Submillimeter Array}

\newpage
Fig. 1: False color 24 $\mu$m image from the Spitzer MIPS for the IRDC G30.88
region. The color bar indicates the logarithmic flux scale 
in units of MJy~sr$^{-1}$. The thin white contours represent the \nh3 (1,1) emission 
from the VLA. The thick yellow contour outlines the JCMT 850 $\mu$m continuum emission.
The dashed circle indicates the FWHM primary beam of the \nh3 observations from the VLA. 
The star symbols mark the dust peaks detected with the SMA. The cross symbols mark the 
position of the \h2O masers. 

Fig. 2: The integrated intensity of the \nh3 (1,1), (2,2) and (3,3)
emission from the VLA, the 850 $\mu$m emission from JCMT,
and 870 $\mu$m emission from the SMA. The \nh3 
images are contoured at an interval of 15 mJy~beam$^{-1}$ $\times$ km s$^{-1}$. 
The JCMT data
are plotted at every 10\% of the peak (2 Jy~beam$^{-1}$). The SMA data are contoured
in steps of 25 mJy~beam$^{-1}$. The star
symbols mark the dust peaks detected with the SMA. The cross symbols mark the position
of the \h2O masers. The dashed circle indicates the FWHM primary beam of the VLA \nh3 
observations. The thin dashed line outlines the 100\% sensitivity of the SMA 
observations. The spatial resolution of each
dataset is marked by the shaded ellipse at the bottom-left corner of each panel.

Fig. 3: Spectra of the \nh3 (J, K) = (1,1), (2,2) and (3,3) from the VLA,
and the \nh3 (3,3) transition from the GBT toward the position of the SMM1.
The two velocity components are marked as ``C1'' and ``C2''.

Fig. 4: Moment 0, 1 and 2 maps of the \nh3 (2,2) line from the VLA for the
two velocity components at 97 \kms-1 (C1), and 107 \kms-1 (C2) respectively.
The contours for the moment 0 map starts at 4  mJy~beam$^{-1}$ $\times$ km s$^{-1}$
and in increments of the same value.
The star symbols mark the dust peaks detected with the SMA. The cross symbols mark the 
position of the \h2O masers. The synthesized beam is marked at the lower-left corner of
each panel.

Fig. 5: \h2O maser spectra for maser features {\it 1, 2, 3, 4, 5, 6,} and {\it 7}.

Fig. 6: \nh3 (1,1) emission in different velocity channels for the C1 (97 \kms-1)
component (Fig. 6a) and C2 (107 \kms-1) component (Fig. 6b). The contour levels
are in steps of 8 mJy~beam$^{-1}$ starting from 8 mJy~beam$^{-1}$.
The star symbols mark the dust peaks detected with the SMA. The cross symbols mark the 
position of the \h2O masers. The synthesized beam is marked at the lower-left corner of
the first panel. The corresponding LSR velocity of the channel is indicated at the top-right
corner of each panel.

Fig. 7: Comparison between the SMA continuum image and simulated observations 
using the JCMT data as the model.
Fig. 7a presents the SMA continuum image at 850 $\mu$m from a single pointing centered
at RA(J2000) = 18:47:12.70, Dec(J2000) = $-$1:45:22.90. Fig. 7b presents
the source model based on the JCMT data tapered by the 34$''$ FWHM primary beam response of the 
SMA. The contour levels in Fig. 7b are plotted at every 10\% of the peak flux of 1.37 Jy per 14$''$
beam. Fig. 7c presents the simulated image of the model dust emission in 7b using 
the UV coverage of the SMA observations. Fig. 7d presents the simulated image of the 
model in Fig. 7b plus a point source of 30 mJy at ($5'', 5''$) offset from the pointing center.
The contour levels in Figs. 7a,c,d are plotted $\pm$17 and $\pm$23  mJy~beam$^{-1}$. The
synthesized beam in the simulated images is $1''9 \times 1''.8$, same as the SMA image in Figure 2.

\newpage

\begin{table*}
\caption{List of Observational Parameters}
\label{tab:1}
\begin{center}
\begin{tabular}{l|lllllcc}
\hline \hline
Instrument & Date of      & Line        &  Bandwidth & Spectral    & rms & Angular \\
           & Observations &             &     (MHz)  & Res. (\kms-1) & (mJy) & Res.  ($''$)   \\
\hline
VLA-D    & 2010/01/08   & \nh3 (1,1),(2,2) &   3.12  &  0.64   &  2.5  & $4 \times 3$      \\
VLA-D    & 2010/05/09   & \nh3 (3,3)       &   4     &  0.20   &  2.5  & $4 \times 3$      \\
VLA-C    & 2010/11/24   & H$_2$O           &   4     &  0.84   &  1.5  & $2 \times 1$   \\  \hline
GBT     & 2010/02/27   & \nh3 (1,1),(2,2)  &   8     &  0.025  &  20  & 30          \\
        & $-$          & (3,3),(4,4)       &        &              \\ \hline
SMA Compact    & 2007/07/07   & $-$        & 4000 &  $-$     &  5.7$^a$  & $1.9 \times 1.8$   \\
SMA Compact    & 2007/10/19   & $-$        & 4000 &  $-$     &  &    \\
SMA Compact    & 2008/06/02   & $-$        & 4000 &  $-$     &  &     \\
SMA Compact    & 2008/10/15   & $-$        & 4000 &  $-$     &  &      \\
\hline
\end{tabular}
\tablenotetext{a}{$1\sigma$ rms in the combined SMA image.}
\end{center}
\end{table*}

\begin{table}[h]
\caption{Physical Parameters of Cores}
\begin{center}
\begin{tabular}{rlllccrrr} \hline \hline
Name & R.A.(J2000)   & Dec.(J2000)  & Flux$^a$ & Mass$^b$ & Association$^c$   \\
     &  ($^h$ $^m$ $^s$)   &  ($^\circ$ $'$ $''$) & (Jy) & (\msun)  &  \\ \hline
SMM1  & 18:47:13.68  & -01:45:03.6 & 0.26 & 32 & C2   \\        %0.3496E+24 cm-2   0.1622E+07 cm-3
SMM2  & 18:47:12.61  & -01:44:42.1 & 0.10  & 40 & C2  \\ %0.1313E+24 cm-2   0.6775E+06 cm-3
SMM3 & 18:47:12.56  & -01:44:52.9 & 0.082 &  33 & C2    \\  % 0.1182E+24 cm-2  0.6098E+06 cm-3
SMM4 & 18:47:12.78  & -01:44.27.4 & 0.067 &  27 & C2    \\  % 0.1257E+24 cm-2     0.6485E+06 cm-3
SMM5 & 18:47:10.32  & -01:45:51.7 & 0.060 &  19 & C1    \\ % 0.1126E+24 cm-2   0.5808E+06 cm-3
SMM6 & 18:47:15.14  & -01:44:46.0 & 0.044 &  18 & C2    \\   \hline        %0.9653E+23 cm-2   0.4743E+06 cm-3
\end{tabular}
\tablenotetext{a}{Integrated flux.}
\tablenotetext{b}{We adopt dust temperatures of 45 K for SMM1, and 19 K for SMM2 through SMM6.
We use a dust emissivity index $\beta$ of 1.5, and the dust opacity law of \citet{hildebrand1983}.
The source distances used are 6.5 kpc for the C1 clump and 7.3 kpc for the C2 clump, respectively. }
\tablenotetext{c}{Association of dust continuum sources with the two cloud components.}
\end{center}
\end{table}

\begin{table}[h]
\caption{Properties of \h2O Masers}
\begin{center}
\begin{tabular}{cllrc} \hline \hline
Name & R.A.(J2000)   & Dec.(J2000)  & Flux$^a$    & Association$^b$ \\
     &  ($^h$ $^m$ $^s$)   &  ($^\circ$ $'$ $''$) &  (K) &  \\ \hline
1  & 18:47:13.70 & -01:45:03.2   & 215.8  & C2 \\	% (-0.133, 0.593)
2  & 18:47:09.07 & -01:44:11.7   & 175.4  & C2 \\        % (-69.38, 52.22)
3  & 18:47:12.67 & -01:44:15.7   & 128.3  & C2 \\	% (-15.564, 47.907)
4  & 18:47:10.30 & -01:45:52.2   & 64.0  & C1 \\	% (-50.935, -48.375)
5  & 18:47:14.67 & -01:43:58.2   & 17.4  & C1 \\	% (14.633, 65.561)
6  & 18:47:15.50 & -01:44:17.2   & 16.8  & C2  \\	% (26.958, 46.351)
7  & 18:47:15.74 & -01:44:45.2   & 14.4  & C2 \\	% (30.148, 18.254)
% 8  & 18:47:14.40 & -01:44:56.2   &  - \\	%
\hline 
\end{tabular}
\tablenotetext{a}{To avoid introducing additional error,
brightness temperatures reported here are not corrected for the
primary beam attenuation.  Assuming a Gaussian primary beam 
with a FWHM of 120$''$, we find a peak brightness temperature of 808.6 K for maser feature 2,
a maser knot outside the FWHM of the primary beam. The Jansky to Kelvin conversion factor
is $1.11 \times 10^3$ Jy~K$^{-1}$. }
\tablenotetext{b}{Association of the \h2O masers with the two cloud components.}
\end{center}
\end{table}

\pagebreak

\begin{figure}[h]
\includegraphics{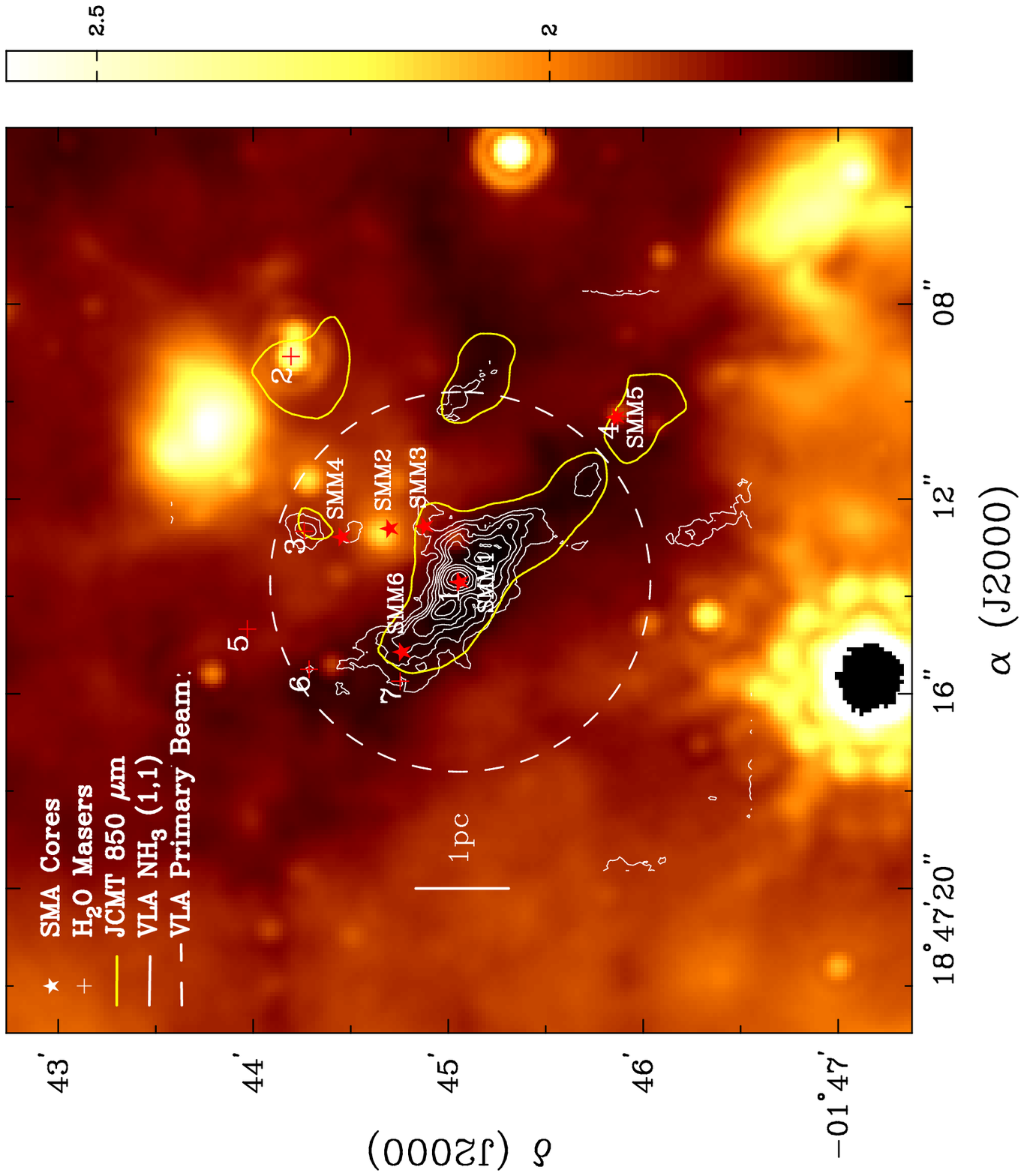}
\caption{ }
\end{figure}

\newpage
\begin{figure}[h]
\includegraphics{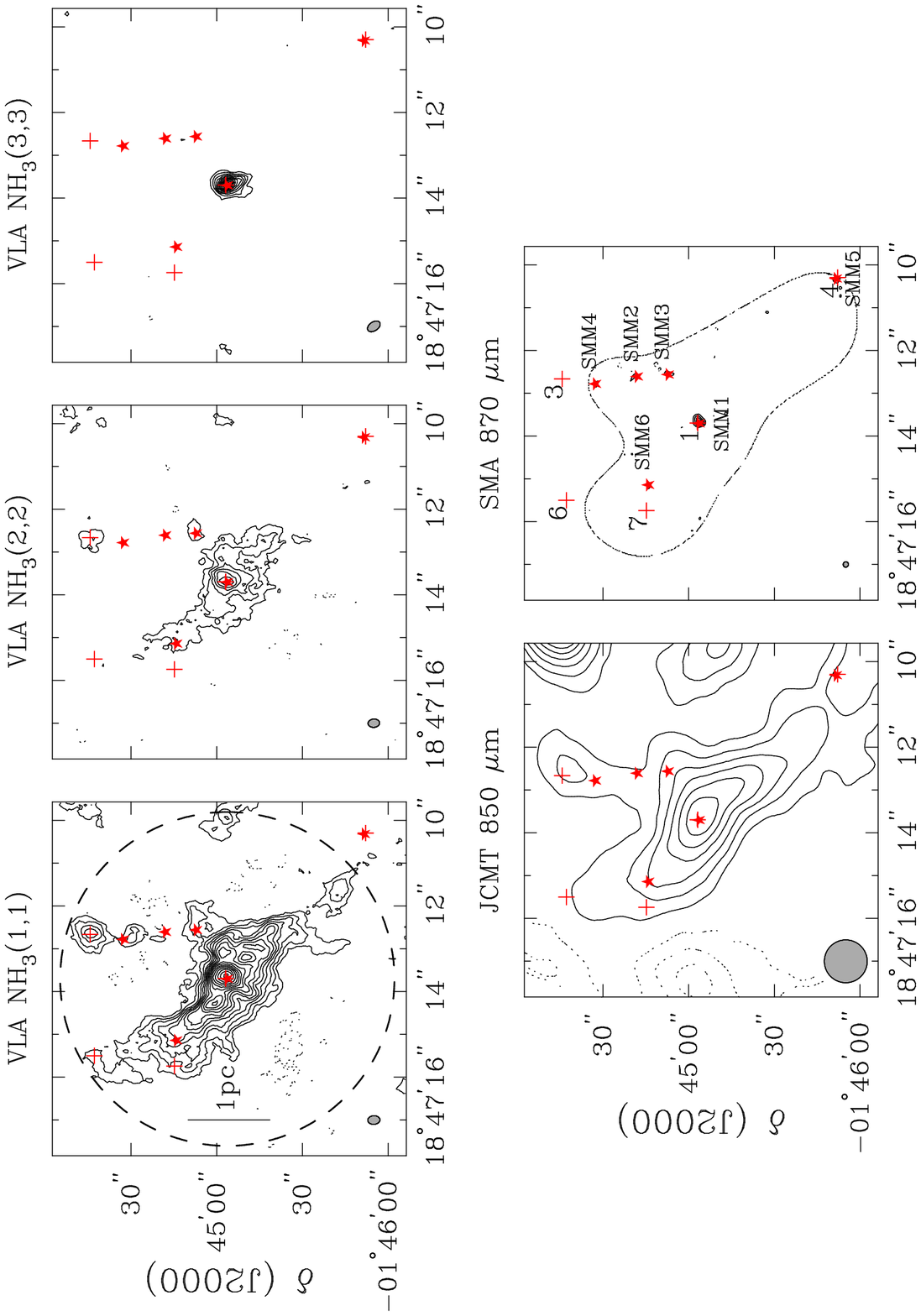}
\caption{ }
\end{figure}

\newpage
\begin{figure}[h]
\includegraphics{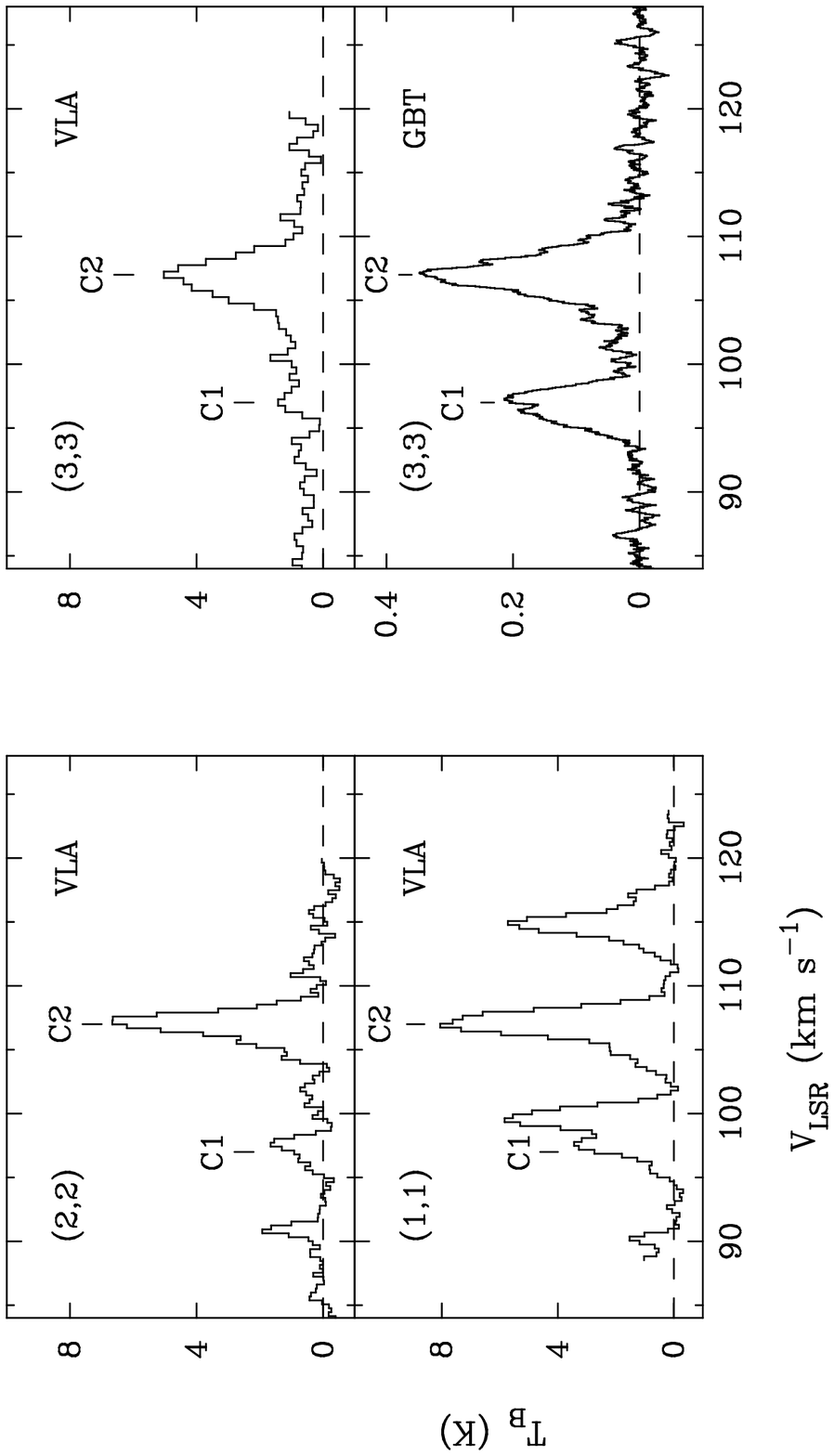}
\caption{ }
\end{figure}

\newpage
\begin{figure}[h]
\includegraphics{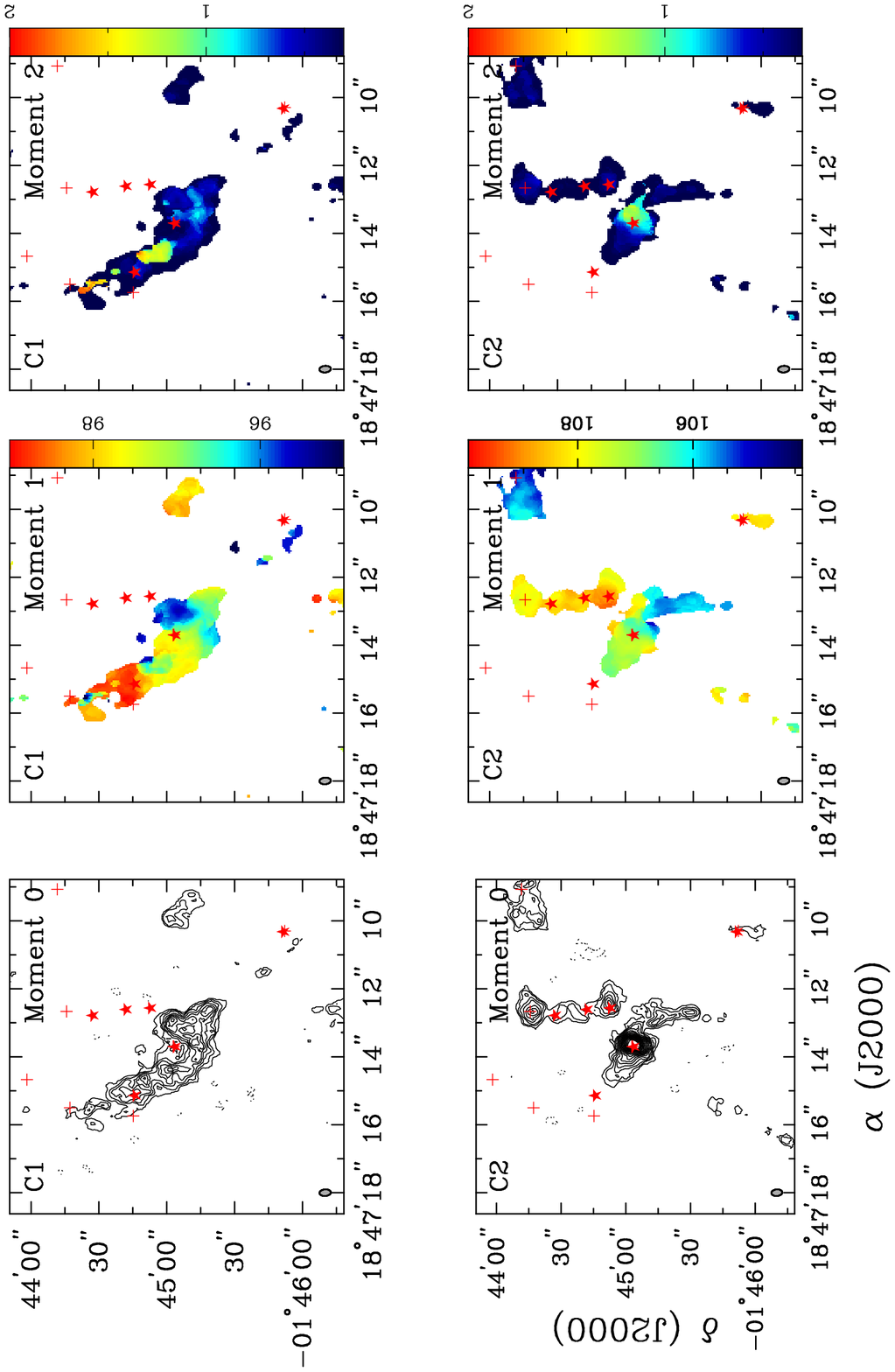}
\caption{ }
\end{figure}

\newpage

\begin{figure}[h]
\includegraphics{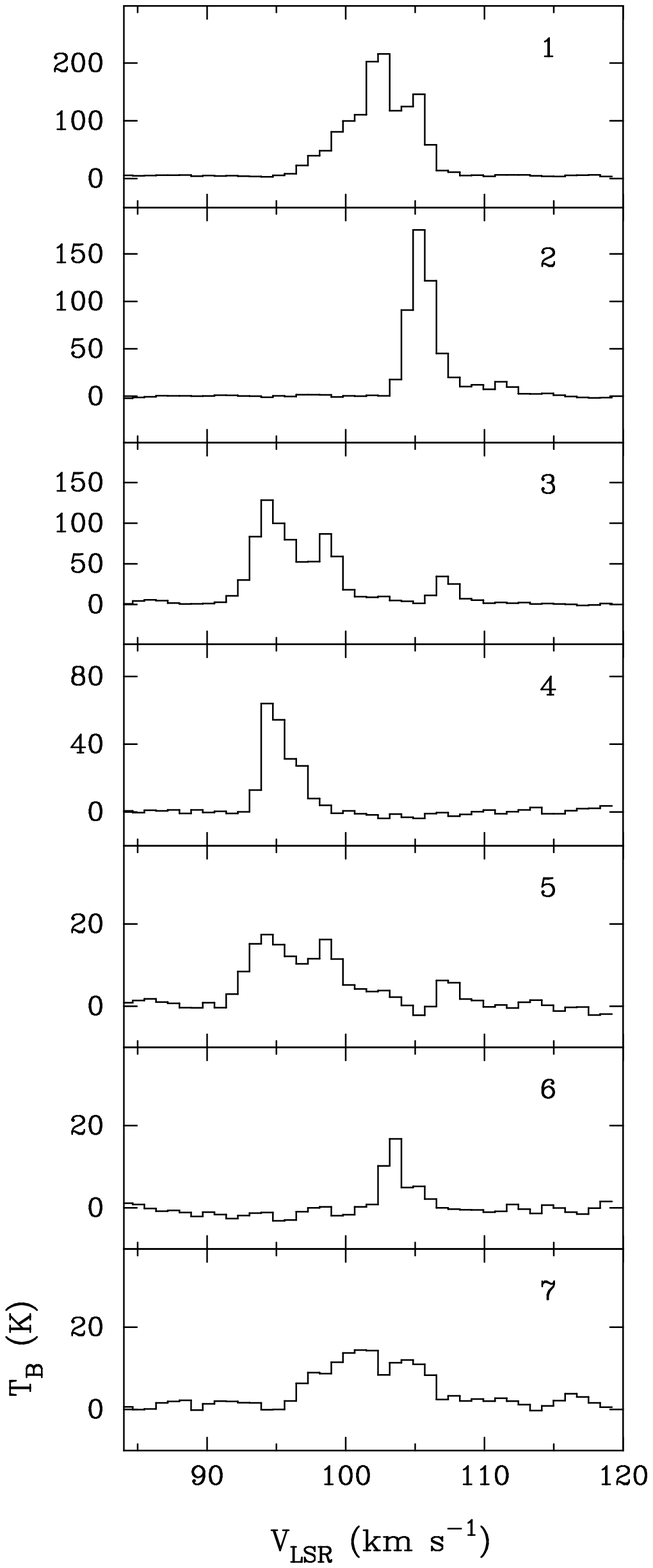}
\caption{ }
\end{figure}
\newpage

\newpage

\begin{figure}[h]
\figurenum{6a}
\includegraphics{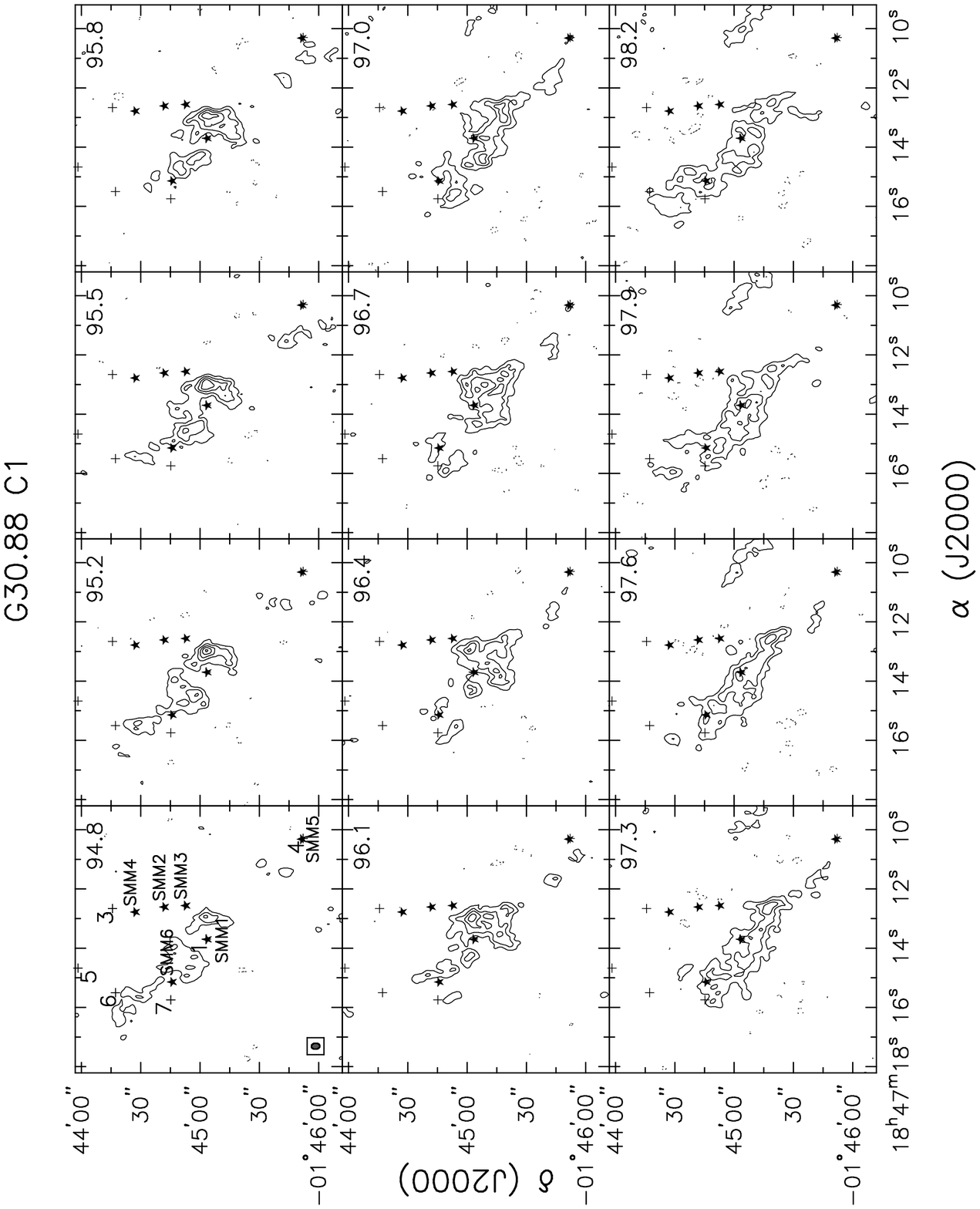}
\caption{ }
\end{figure}
\newpage

\newpage

\begin{figure}[h]
\figurenum{6b}
\includegraphics{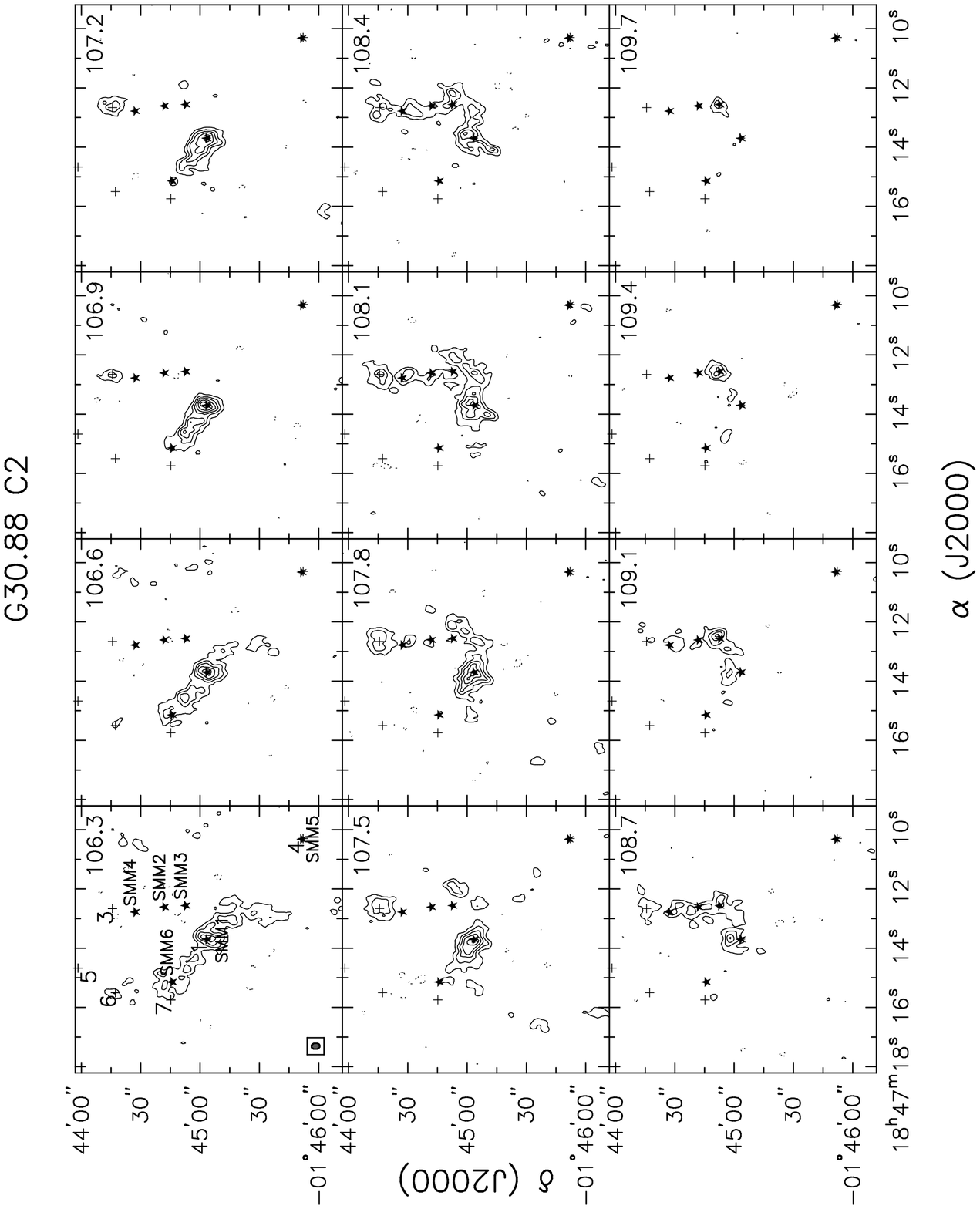}
\caption{ }
\end{figure}
\newpage

\newpage

\begin{figure}[h]
\figurenum{7}
\includegraphics{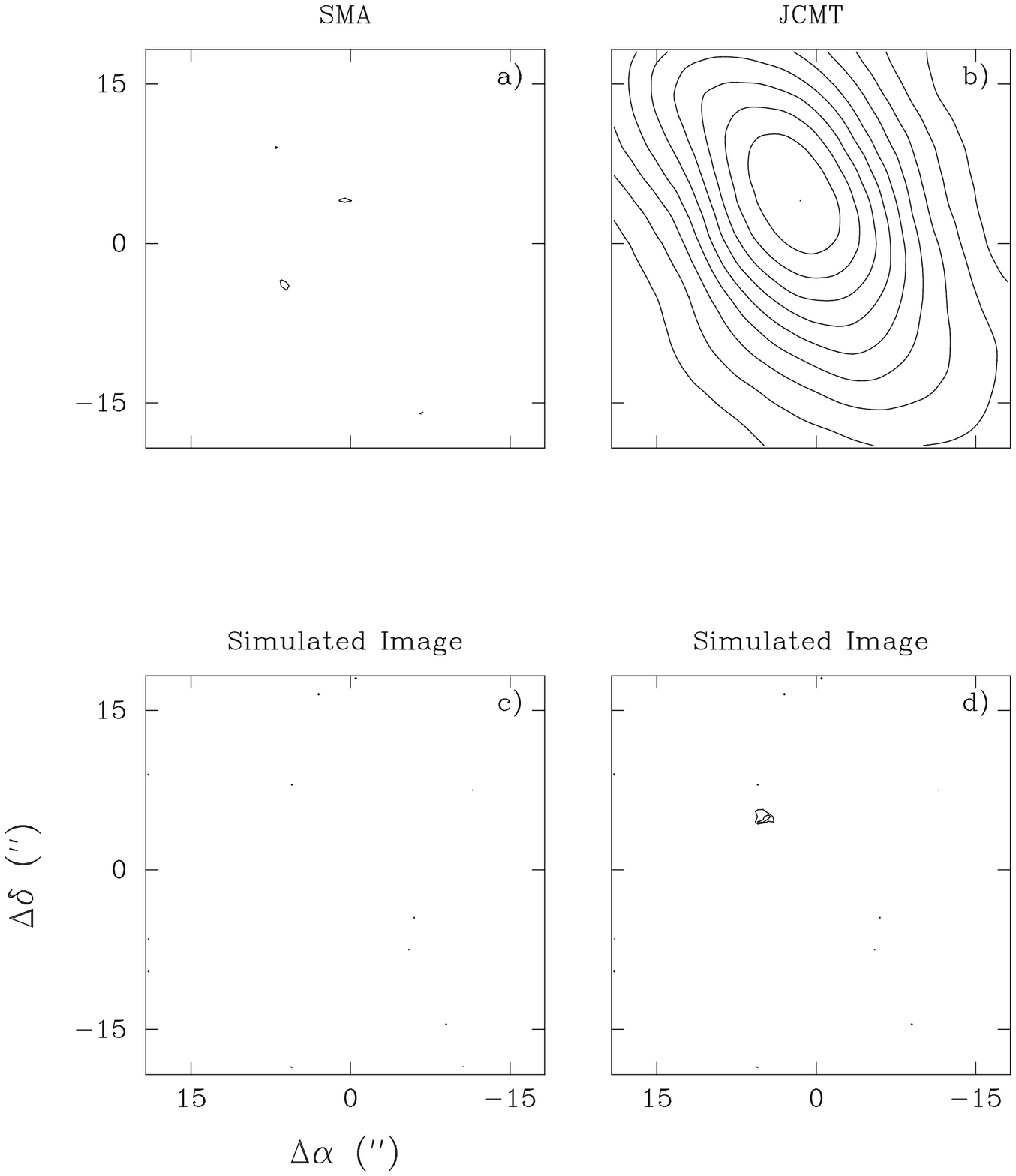}
\caption{ }
\end{figure}

%\newpage
%\input{ms_zhang_rev.bbl}
%\input{/home/qzhang/tex/acronyms}
%\bibliography{/home/qzhang/tex/bibliography}
%\bibliographystyle{/home/qzhang/tex/aa} % this does the style, aa.bst

\end{document}